\documentclass[12pt]{article}
\usepackage{amssymb,amsmath,cite}
\catcode`\@=11

\@addtoreset{equation}{section}
\def\theequation{\arabic{section}.\arabic{equation}}
     
\catcode`\@=11
\def\thesection{\arabic{section}.}

\def\appendix{\setcounter{section}{0}
        \def\thesection{Appendix.}
        \def\theequation{\Alph{section}.\arabic{equation}}}
\def\section{\@startsection{section}{1}{\z@}{3.5ex plus 1ex minus
   .2ex}{2.3ex plus .2ex}{\large\bf}}
     
\long\def\@makefntext#1{\parindent 0cm\noindent
\hbox to 1em{\hss$^{\@thefnmark}$}#1}

\newcommand{\captionfonts}{\small}
\makeatletter  
\long\def\@makecaption#1#2{%
  \vskip\abovecaptionskip
  \sbox\@tempboxa{{\captionfonts #1: #2}}%
  \ifdim \wd\@tempboxa >\hsize
    {\captionfonts #1: #2\par}
  \else
    \hbox to\hsize{\hfil\box\@tempboxa\hfil}%
  \fi
  \vskip\belowcaptionskip}
\makeatother   
 
\begin{document}
\begin{titlepage}
\begin{flushright}
February 2008\\
\end{flushright}
\vspace{.5in}
\begin{center}
{\Large\bf
 Is Quantum Gravity Necessary?}\\
\vspace{.4in}
{S.~C{\sc arlip}\footnote{\it email: carlip@physics.ucdavis.edu}\\
       {\small\it Department of Physics}\\
       {\small\it University of California}\\
       {\small\it Davis, CA 95616}\\{\small\it USA}}
\end{center}

\vspace{.5in}
\begin{center}
{\large\bf Abstract}
\end{center}
\begin{center}
\begin{minipage}{4.75in}
{\small 
In view of the enormous difficulties we seem to face in  
quantizing general relativity, we should perhaps consider 
the possibility that gravity is a fundamentally classical 
interaction.  Theoretical arguments against such mixed
classical-quantum models are strong, but not conclusive, 
and the question is ultimately one for experiment.  I review 
some work in progress on the possibility of experimental 
tests, exploiting the nonlinearity of the classical-quantum 
coupling, that could help settle this question.
}
\end{minipage}
\end{center}
\end{titlepage}
\addtocounter{footnote}{-1}

The first attempts to quantize general relativity date back to
the early 1930s \cite{Rovelli}.  In the 75 years that have
followed, we have learned an enormous amount: gauge-fixing
and Faddeev-Popov ghosts, background field methods, the effective 
action formalism, the canonical analysis of constrained systems, 
the investigation of gauge-invariant observables, and much of what 
we know about topology in quantum field theory grew out of attempts 
to quantize gravity.  But despite the extraordinary work of a great 
many outstanding physicists, a complete, consistent, and compelling 
quantum theory of gravity still seems distant \cite{Carlip}.  

In view of this history, we should perhaps consider the possibility
that we are asking the wrong question.  It could be that gravity 
is simply not quantum mechanical.  The prospect of a fundamentally 
classical theory of gravity is unpalatable; in Duff's words 
\cite{Duff}, it ``seems to be the very antithesis of the economy 
of thought which is surely the basis of theoretical physics.''  But 
the matter is ultimately one for experiment.  As Rosenfeld has put 
it \cite{Rosenfeld},
\begin{quote}
It is nice to have at one's disposal such exquisite mathematical
tools as the present methods of quantum field theory, but one 
should not forget that these methods have been elaborated in order 
to describe definite empirical situations, in which they find their 
only justification.  Any question as to their range of application 
can only be answered by experience, not by formal argumentation.  
Even the legendary Chicago machine cannot deliver the sausages if 
it is not supplied with hogs.
\end{quote}

There are old arguments that fundamentally classical fields are 
incompatible with quantum mechanics, in the sense that they could 
be used to violate the uncertainty principle \cite{DeWitt}.  Details 
depend on how the classical field interacts with a quantum system.  
Eppley and Hannah \cite{Eppley,Callender} have considered two cases:
\begin{enumerate}
\item A classical gravitational measurement collapses the quantum 
wave function: then momentum is not conserved.  Consider a quantum 
object in a coherent state with a very small uncertainty in momentum
and a correspondingly large uncertainty in position.  Measure its
position by scattering a very short wavelength gravitational wave,
causing its state to change to one with a very small uncertainty in 
position and a large uncertainty in momentum.  If gravity is classical,
the gravitational wave can carry an arbitrarily small momentum, 
despite its short wavelength; yet by the uncertainty principle, the 
quantum system must sometimes experience a large change in momentum.
\item A classical gravitational measurement does not collapses the 
quantum wave function: then signals can be sent faster than light.
Place a proton in a box, in a state in which it has an equal
probability of being in the left or right half.  Split the box in 
half and carry one half to a remote location.  Monitor your half 
continuously with gravitational measurements, while a colleague 
performs a nongravitational measurement of the other half.  Your 
colleague's measurement will collapse the wave function, causing an 
instantaneous and detectable change in the half of the box you are 
monitoring.
\end{enumerate}
Page and Geilker \cite{Page} add a third case:
\begin{enumerate}\addtocounter{enumi}{2}
\item Neither classical nor quantum measurements collapse the wave
function (Everett interpretation): then gravitational fields will
not be observed to have localized sources.  Consider a gravitating 
mass in a superposition of two widely separated position eigenstates.
If its classical gravitational field depends on its quantum wave 
function, its gravitational attraction should point toward some
intermediate ``average'' location \cite{Kibblea,Unruh}.  Page and 
Geilker tested this experimentally, but the outcome is already 
apparent in, say, the observed gravitational field of the Moon.
\end{enumerate}

But while such arguments are certainly suggestive, they are 
not really conclusive \cite{Rosenfeld,vonB,Callender,Mattingly}.  
For instance, there are inherent non-quantum limits to 
gravitational measurements \cite{vonB,Smolin}, whose implications 
for an Eppley-Hannah-type argument have yet to be fully explored.  The 
general question of whether one can consistently couple classical 
and quantum systems is a matter of ongoing research---see, for 
example, \cite{Sudarshan,Jones,Halliwell,Caro,DiosiGisin,Peres,Terno}%
---and is not yet resolved.  

The thought experiments of Eppley, Hannah, and others do, however, 
suggest that a fundamentally classical theory of gravity is likely 
to require changes to quantum mechanics as well.  As I shall argue
below, once one allows a coupling between classical and quantum
systems, quantum mechanics almost inevitably becomes nonlinear,
suggesting the possibility of sensitive new experimental tests.

\section{Semiclassical gravity and the Schr{\"o}dinger-Newton equation}

If we wish to couple classical gravity and quantum matter, we
need field equations for gravity.  The standard Einstein equations,
\begin{equation}
G_{ab} = 8\pi {\hat T}_{ab} ,
\label{a1}
\end{equation}
no longer make sense, since they now equate a c-number with an operator.
We might try to interpret (\ref{a1}) as an eigenvalue equation, but
this picture fails: the components of the stress-energy tensor do 
not commute, and cannot be simultaneously diagonalized \cite{Unruh}.  

The obvious next step is to replace the right-hand side of (\ref{a1}) 
with an expectation value,
\begin{equation}
G_{ab} = 8\pi \langle\psi|{\hat T}_{ab}|\psi\rangle ,
\label{a2}
\end{equation}
leading to the model of ``semiclassical gravity'' first proposed by 
M{\o}ller \cite{Moller} and Rosenfeld \cite{Rosenfeld}, and derived
from an action principle by Kibble and Randjbar-Daemi \cite{Kibble}.
Seen merely as a Hartree approximation to a full quantum theory 
of gravity, such a model seems uncontroversial.  But as Kibble and 
Randjbar-Daemi emphasized \cite{Kibble}, seen as a fundamental 
theory, the model implies nonlinearities in quantum mechanics: 
the Schr{\"o}dinger equation for the wave function $|\psi\rangle$ 
depends on the metric, which now depends in turn on the wave 
function.\footnote{Dirac was also apparently aware of this; see 
\cite{vonB}, p.\ 1.}  Adler has observed that semiclassical gravity 
contains self-interaction terms that are not present in a Hartree 
approximation \cite{Adler}, further differentiating it from a mere 
approximation to a full quantum theory.

Several technical problems with semiclassical gravity have been
pointed out in the literature.  Field redefinition ambiguities can 
lead to inequivalent quantizations of the same classical theory 
\cite{Duff}; renormalization may either require classical
curvature-squared terms in the action that can lead to negative
energies \cite{Woodard} or new matter vertices that imply
noncausal behavior at short distances \cite{Anselmi}; and it is not 
obvious that an abrupt change in the right-hand side of (\ref{a2}) due 
to wave function collapse can be consistent with conservation of the 
left-hand side \cite{Unruh}.  Again, though, these objections do not 
seem conclusive.  The nonlinearity of semiclassical gravity, on the 
other hand, suggests that experimental tests may be possible: gravity 
is very weak, but limits on nonlinearities in quantum mechanics are 
very strong \cite{Weinberg}.

To address this question, it is useful to start with the Newtonian
approximation to (\ref{a2}), the Schr{\"o}dinger-Newton equation
\cite{Diosi,Penrose}
\begin{equation}
i\hbar\frac{\partial\psi}{\partial t} =
-\frac{\hbar^2}{2m}{\nabla}^2\psi - m\Phi\psi ,\qquad
{\nabla}^2 \Phi = 4\pi G m|\psi|^2 .
\label{a3}
\end{equation}
As in full semiclassical gravity, this model treats matter quantum
mechanically, but describes gravity in terms of a classical Newtonian
potential $\Phi$ sourced by the expectation value of the mass density.
Despite the nonlinearities of the coupled system (\ref{a3}), the 
standard probability interpretation of the wave function remains 
consistent; in particular, the probability current continuity equation
\begin{equation}
 \frac{\partial}{\partial t}|\psi|^2 
  = \vec{\nabla}\cdot\left[ \frac{i\hbar}{2m}\left( 
  \psi^*\vec{\nabla}\psi - \psi\vec{\nabla}\psi^*\right)\right] 
\label{a4}
\end{equation}
still holds, and total probability is conserved.  A number of authors 
have studied this system \cite{Moroz,Bernstein,Tod,Todb}, and we know 
a good bit about the stationary states with low energy eigenvalues, 
but time evolution has proven to be much more problematic 
\cite{Harrison,Harrisonb,Guzman}.

\section{Estimates and numerics}

The question, then, is whether the nonlinearities in the 
Schr{\"o}dinger-Newton equation (\ref{a3}) are large enough to 
lead to observable consequences.  Let us begin with a rough estimate.
Consider a particle of mass $m$ with a localized initial wave function%
---for simplicity, a Gaussian,
\begin{equation}
\psi(r,0) = \left( \frac{\alpha}{\pi} \right)^{3/4} e^{-\alpha r^2 /2}
\label{b1}
\end{equation}
with width $\alpha^{-1/2}$.  The time evolution of $\psi$ will depend 
on two competing effects, the quantum mechanical spreading of the
wave function and its Newtonian ``self-gravitation,'' the latter
arising because semiclassical gravity treats a wave function as a 
distributed source.  For a very low mass, self-gravitation should
be negligible, while for a high enough mass, the wave function should
undergo ``gravitational collapse.''

To estimate the critical mass at the boundary between wave packet 
spreading and collapse, note first that the peak probability density 
for a free particle occurs at 
\begin{equation}
r_p \sim \alpha^{-1/2}\left(1+\frac{\alpha^2\hbar^2}{m^2}t^2\right)^{1/2},
\label{b2}
\end{equation}
which ``accelerates'' outward at a rate $a_{\mathit{out}} = %
{\Ddot r}_p \sim \hbar^2/m^2r_p{}^3$.  This will balance the inward
gravitational acceleration $a_{\mathit{in}} \sim Gm/r_p{}^2$ at $t=0$ 
when
\begin{equation}
m \sim \left(\frac{\hbar^2\sqrt{\alpha}}{G}\right)^{1/3}  .
\label{b3}
\end{equation}
This is almost certainly an overestimate: as $t$ increases, 
$a_{\mathit{out}}$ drops more quickly than $a_{\mathit{in}}$, so
even if wave packet spreading dominates initially, self-gravity
may eventually win.

For more precise results, one must solve (\ref{a3}) numerically.  Note 
that although the initial data (\ref{b1}) depend on two parameters,  
$\alpha$ and $m$, the Schr{\"o}dinger-Newton equation is invariant under 
the rescalings
\begin{equation}
m\rightarrow \mu m ,\qquad \vec{x}\rightarrow \mu^{-3}\vec{x},\qquad 
t\rightarrow \mu^5t ,\qquad \psi\rightarrow \mu^{9/2}\psi ,
\label{b4}
\end{equation}
so it is enough to consider a one-parameter set of solutions.
Peter Salzman and I have numerically simulated the evolution of an 
initial Gaussian wave function \cite{Carlipa,Salzman}.  We find the 
expected qualitative results:
\begin{enumerate}
\item For small masses, the behavior is virtually identical to that 
  of a free particle, while as $m$ increases, the wave packet spreads 
  more slowly.
\item In a transitional range of mass, the wave packet is unstable,
  fluctuating rapidly and developing growing oscillations.  (A similar
  instability is seen in \cite{Harrison,Harrisonb,Guzman}.)
\item For large masses, the wave packet undergoes ``gravitational
  collapse.'' 
\end{enumerate}
Surprisingly, though, we find that the ``collapse'' behavior occurs
at considerably lower masses than the estimate (\ref{b3}) suggests.
For the initial width of $\alpha = 5\times10^{16}\,\mathrm{m}^{-2}$
used in the simulations, the mass (\ref{b3}) is on the order of
$10^{10}\,\mathrm{u}$, while collapse first appears in the simulations
for masses of about $10^4\,\mathrm{u}$.\footnote{Anticipating a 
discussion of molecular interferometry, I am giving masses in unified 
atomic mass units.}  This result is somewhat unexpected, although not 
implausible in view of the highly nonlinear nature of the problem.  
Fortunately, it is now being tested by another group, using different, 
independently developed code. 

Assuming the validity of our simulations, we can use the scaling 
behavior (\ref{b4}) to obtain the parameters for gravitational 
collapse.  We find that a wave packet of initial width $w=\alpha^{-1/2}$ 
will shrink if its initial mass lies in a range $m_-(w)<m<m_+(w)$, 
with 
\begin{equation}
m_-/1\,\mathrm{u} = 1300({w/1\,\mathrm{\mu m}})^{-1/3}, 
  \qquad
m_+/1\,\mathrm{u} = 4.8\times10^{13}({w/1\,\mathrm{\mu m}})^{-1/3} .
\label{b6}
\end{equation}
(For $m>m_+$, we have not been able to run the simulation long enough 
to reliably determine the outcome.)
The numerically obtained collapse times, in nanoseconds, are
\begin{equation}
t_-/1\,\mathrm{ns} = 1.2\times10^{-4}({w/1\,\mathrm{\mu m}})^{-5/3}
  \qquad
t_+/1\,\mathrm{ns} = 1.2\times10^{-2}({w/1\,\mathrm{\mu m}})^{-5/3} .
\label{b7}
\end{equation}

\section{Experimental tests}

Are nonlinearities at the level described above experimentally 
testable?  To get a measurable signal, one needs to use as large a 
mass as possible while still maintaining observable quantum behavior.  
The best bet seems to be molecular interferometry, where a ``collapsing'' 
wave packet would lead to suppression of interference.  At this
writing, the heaviest molecule that has experimentally exhibited 
interference is fluorofullerene, C${}_{60}$F${}_{48}$, with a mass 
of $1632\,\mathrm{u}$ \cite{Hackermuller}.  The grating slits in 
the fluorofullerene experiment have a width $w\sim.5\,\mathrm{\mu m}$.  
From (\ref{b6}), semiclassical gravity would predict a loss of 
interference for a wave packet of this width for masses greater 
than about $m\sim 1600\,\mathrm{u}$.  Fluorofullerene lies just at 
the edge of this range.  

Unfortunately, this is too optimistic an estimate: the wave packets
in molecular interferometry experiments are not spherically symmetric
Gaussians, and their effective width may be quite a bit larger.  In
\cite{Carlipa}, we estimate that the fluorofullerene experiment may
be fall short of a real test by a factor of about 500.  

This leaves work for both experimentalists and theorists.  On the 
theory side, assuming the results of \cite{Carlipa,Salzman} are 
confirmed, we need to look at more realistic initial profiles.
It will be important to see how sensitive the ``collapse'' is to
the shape of the initial wave function---we cannot yet rule out
the possibility that the behavior we see is an artifact of the
Gaussian initial conditions---and to see how the collapse time 
depends on the wave packet shape.

On the experimental side, some progress can come from reducing
and better controlling the wave packet width, for example by using
shutters to limit the longitudinal extent of the packet.  The 
most important gain, though, will come from the move to heavier 
molecules.  A number of experimentalists have predicted that with 
improved methods---optical gratings, for example---it should be 
possible to observe interference for molecules with masses as high 
as $10^6\,\mathrm{u}$ \cite{Zeilinger,Talbot,Hornberger,Zeilingerb}.  
If the next generation of molecular interferometry experiments can 
come even close to this limit, a clean test of semiclassical gravity 
should be well within reach.

\vspace{1.5ex}
\begin{flushleft}
\large\bf Acknowledgments
\end{flushleft}

I would like to thank my colleagues at Peyresq 11, including Brandon
Carter, Larry Ford, Bei-Lok Hu, Seif Randjbar-Daemi, and Albert Roura, 
for a great many useful comments and suggestions.  This work was supported 
in part by U.S.\ Department of Energy grant DE-FG02-91ER40674.

\end{document}